\documentstyle[twocolumn,aps,epsf]{revtex}

\def\myfigure#1#2{{\leftskip=0.000753\textwidth \rightskip\leftskip\small
\begin{figure}\baselineskip=14pt plus 2pt minus 1pt
\centerline{#1}\nobreak\smallskip\nobreak #2\end{figure}}}
\global\firstfigfalse

\draft

\begin{document}

\title {A Novel Monte Carlo Approach to the Dynamics of Fluids
--- Single Particle Diffusion, Correlation Functions and Phase Ordering of Binary Fluids
}
\author{P.\ B.\ Sunil Kumar and Madan Rao}

\address{Institute of Mathematical Sciences, Taramani, Madras 600 113, India}

\date{\today}

\maketitle

\begin{abstract}
We propose a new Monte Carlo scheme to study the late-time dynamics of a 
2-dim hard sphere fluid, modeled by a 
tethered network of hard spheres. Fluidity is simulated by 
breaking and reattaching the flexible tethers. We study 
the diffusion of a tagged particle, and show that the velocity autocorrelation
function has a long-time $t^{-1}$ tail. We investigate the dynamics of
phase separation of a binary fluid at late times, and show that
the domain size $R(t)$ grows as $t^{1/2}$ for high viscosity
fluids with a crossover to $t^{2/3}$ for low viscosity fluids. Our scheme can accomodate particles interacting with a pair potential $V(r)$, and
 modified to study dynamics of fluids in three dimensions.
\end{abstract}

\bigskip


With the advent of modern computers, there have been several large
scale molecular dynamics (MD) \cite{MD}, 
lattice-gas automata (LG) \cite{LGA}
and Langevin simulations (LS) \cite{REVIEW} of the late-time dynamics of fluids.
These simulations have been used to compute dynamical correlation
functions and to study fluid flow in 
different geometries. MD and LS  have also been used to
 study the dynamics of phase ordering of binary
fluids, where the relative concentration of the fluids is coupled to the
fluid velocity. In many such applications, one is interested in late-time
hydrodynamical behaviour. Since MD provides an accurate description of the
microscopic physics, it may not probe dynamics at such late time, unless one makes a sizeable computational investment. LS are a more
coarse-grained description\,; however the dynamical equations are too complicated to solve, and it is not clear that the various simplifying
approximations made, do not miss out features which might affect the physics at large length and time scales.

There is a considerable advantage if Monte Carlo (MC) simulations could
be used instead. MC simulations \cite{MC} have been remarkably successful in the study of the dynamics of alloys, magnets and so on, but
have never been used in dynamical studies involving fluids. 
This is because
there is no natural way to incorporate the momentum density in a MC simulation. 
It would be desirable to devise an MC scheme to
handle fluidity, since it is more coarse-grained than MD and more microscopic
than LS. More recently Lattice Boltzmann (LB)\cite{LB} simulations have 
been employed with this specific goal in mind, but since such simulations 
do not depend on the specific form of the hamiltonian, one has to choose
the equilibrium distribution functions consistent with
interfacial profiles and equilibrium densities.

For the stated reasons and more, we present in this Letter,
a novel MC technique which successfully simulates the fluidity of a fluid.
Though the algorithm can be extended to dimensions $d \ge 2$, and
to any pair potential, we demonstrate its
efficacy for a hard sphere fluid in $d=2$. We determine the single-particle diffusion coefficient and the long-time tail of the
velocity autocorrelation function for dense fluids. We also study
the kinetics of phase ordering of a binary fluid in 2-d and
compute the concentration correlation function to extract the growth laws
described below.

Our model 2-d fluid consists of hard spherical beads (vertices) (the total number of beads $N$, is fixed), restricted onto a finite region of area
$A$ of ${\cal R}^2$. All the
beads are linked together by straight flexible tethers (bonds), in such a way as to triangulate this region. The tethers do not intersect
each other --- each configuration is thus a {\it connected planar graph}.
We ensure that the local coordination number of every bead lies between 3 and 9.
Since the particles are restricted to a 2-d plane with zero curvature,
the distribution of local coordination numbers should be symmetric about 6.
The beads are infinitely
repulsive at distances less than a bead diameter $l_{min}=a$ (for all vertex pairs) or greater than $l_{max}$ (for vertices connected by tethers), and so
the local tether length can vary between these two limits. Our choice of 
$l_{max} = 5 \sqrt{3} a$, guarantees that half of the attempted MC moves are accepted. We distinguish between external (edge) vertices ($V^{E}$) 
and internal (bulk) vertices ($V^{I}$). An external (edge) tether
($T^{E}$) connects two external vertices, while an internal (bulk) tether ($T^{I}$) connects at least one internal vertex. 

Our Monte Carlo simulation should allow 
for configuration changes which sample {\it all 
permissable} regions in phase space.
The configuration changes of our 2-d fluid  consist of the movement of beads with fixed connectivity ({\it bead moves})
and the movement of tethers with fixed vertex positions ({\it flip moves}). 
As we shall see below, the flip moves are crucial in maintaining fluidity.
We shall use rigid boundary conditions (on a hexagonal frame), and so
neither $V^{E}$ nor $T^{E}$ are moved. 
 
The bead moves are effected by
randomly choosing a bead $i$ and then 
translating $i$ to a random point (with a uniform distribution)
within a square (of size $l=2a$)
centered on the old position of $i$. The movement is accepted 
if the new bond lengths lie between $l_{min}$ and $l_{max}$ and 
{\it the graph remains planar} (i.e., no bond intersections).   With each
particle $i$, we can identify a unique n-gon (concave or convex) whose vertices
$\{v_1, v_2, \ldots, v_n\}$, arranged in cyclic order, are connected to $i$.
Let us denote the areas of the triangles $\{iv_1v_2, \ldots, iv_{n-1}v_n\}$ as $\{A(iv_1v_2), \ldots, A(iv_{n-1}v_n)\}$.
Planarity is maintained if $A(iv_1v_2) + \ldots + A(iv_{n-1}v_n)$ 
remains unchanged after the move on $i$. Our unit of time is set by
one Monte Carlo sweep (MCS), defined as $N$ attempted bead moves. 

Fluidity can be generated by a combination of bead and flip moves.
The flip moves are a modification of the bond reconnection algorithm, developed to study the equilibrium behaviour of fluid membranes \cite{MEMB}. A bond $t_{ij}$ (connecting vertices $i$ and $j$) is picked at random. With every internal bond $t_{ij}$, we can identify
two triangles $ijv_1$ and $ijv_2$ on either side of 
$t_{ij}$ which have the {\it smallest area}.
This defines a quadrilateral, with $i$ and $j$ being a pair of 
opposite vertices. The bond $t_{ij}$ is now flipped, only if $v_{1}$ and
$v_2$ are not already connected by a bond, so that it now connects $v_1$ 
and $v_2$ (leaving the vertices $i$ and $j$ unconnected by a bond).
This flip is accepted provided the length of $t_{v_1v_2}$ is less than the maximum allowed length $l_{max}$, and if $t_{v_1v_2}$ {\it does not intersect
any other tether} (planarity). Note that the total number of bonds is conserved 
during this operation. During one MCS, we make $N_{flip}$ 
attempts at flipping bonds.
  
How well does this algorithm mimic the statics and dynamics of a simple fluid ? 
The equilibrium properties of a simple fluid can be derived from the partition
function $Z = Tr_{\bf g} Tr_{\rho} \exp(-\beta F[\{\bf g\},\{\rho\}])$ (where 
$F = \int \,[\,g^2/2\rho_0 + V (\{\rho\})\,]$ is the
free-energy functional of the local momentum density ${\bf g} ({\bf r}, t)$
and the mass density $\rho({\bf r}, t)$ of the fluid).
In our simulation, we define the local velocity for the $i$-th particle, 
${\bf u}_i = {\bf g}_i/m$, as its 
vector displacement in a single MCS. Even though
particle displacements are picked out with a uniform distribution 
from a square of side $l$ centred at the original position of the particle, large displacements, which have a high chance of violating the tethering and 
planarity constraints, are suppressed.
Clearly, the larger the coordination number of the $i$-th particle, the higher is the chance that 
large displacements are suppressed and so the smaller is $\vert {\bf u}_i \vert$. The particle displacements across the whole sample are
independent of each other and so by the central limit theorem,
the displacements (and hence the velocities) of particles tends 
to a normal (Maxwellian) distribution over several MC times. The time 
scale over which this happens is the velocity equilibration time $\tau_{u}$.
From equipartition, a computation of $\langle {u_i}^2 \rangle$ gives the
kinetic temperature $2k_BT/m$ where $m$ is 
the mass of the particle. On the other hand, the trace over $\rho$ is identical to a sum over configurations with (dynamically) varying local coordination number. Since the tethering is dynamical, there is no restriction on the 
allowable configurations sampled by our fluid.
 
Equilibrium is achieved through collisions between particles which result
in velocity exchanges. In our simulations, ``collisions'' occur either through 
the rejection of certain MC moves due to the hard sphere constraint, or through
 flip moves. Collisions resulting from flip moves, transform particles
with low coordination number to high and vice versa. From our discussion above, 
this results in an exchange of velocities. It is clear that particle 
momenta (and angular momenta) are not conserved during an elementary ``collision'', and so our algorithm will not mimic fluid dynamics at 
short time scales. However, averaged over several collisions, both momenta
and angular momenta can be seen to be conserved. Just as in a Langevin
simulation, momentum is conserved only in a statistical sense.
Our MC dynamics thus mimics the dynamics of a fluid over several ``collision''
time scales. Starting with a random triangulation (which we generate by
repeated bead and flip moves on an initial triangular lattice),
 we allow the fluid to
evolve via our MC algorithm. We measure various dynamical transport 
coefficients after discarding all configurations arising
from typically the first 10 Monte Carlo sweeps. 
The single-particle diffusion coefficient $D_{s}$
can be extracted by measuring the mean-square displacement 
of a tagged particle (averaged over several initial conditions and particles)
for different values of $N_{flip}$ and density (Fig.\ 1).
 Not surprisingly, $D_{s}$ 
(and hence the inverse viscosity $\eta^{-1}$) is an
increasing function of $N_{flip}$ (inset Fig.\ 1) at constant density, before it saturates.  

\myfigure{\epsfysize4in\epsfbox{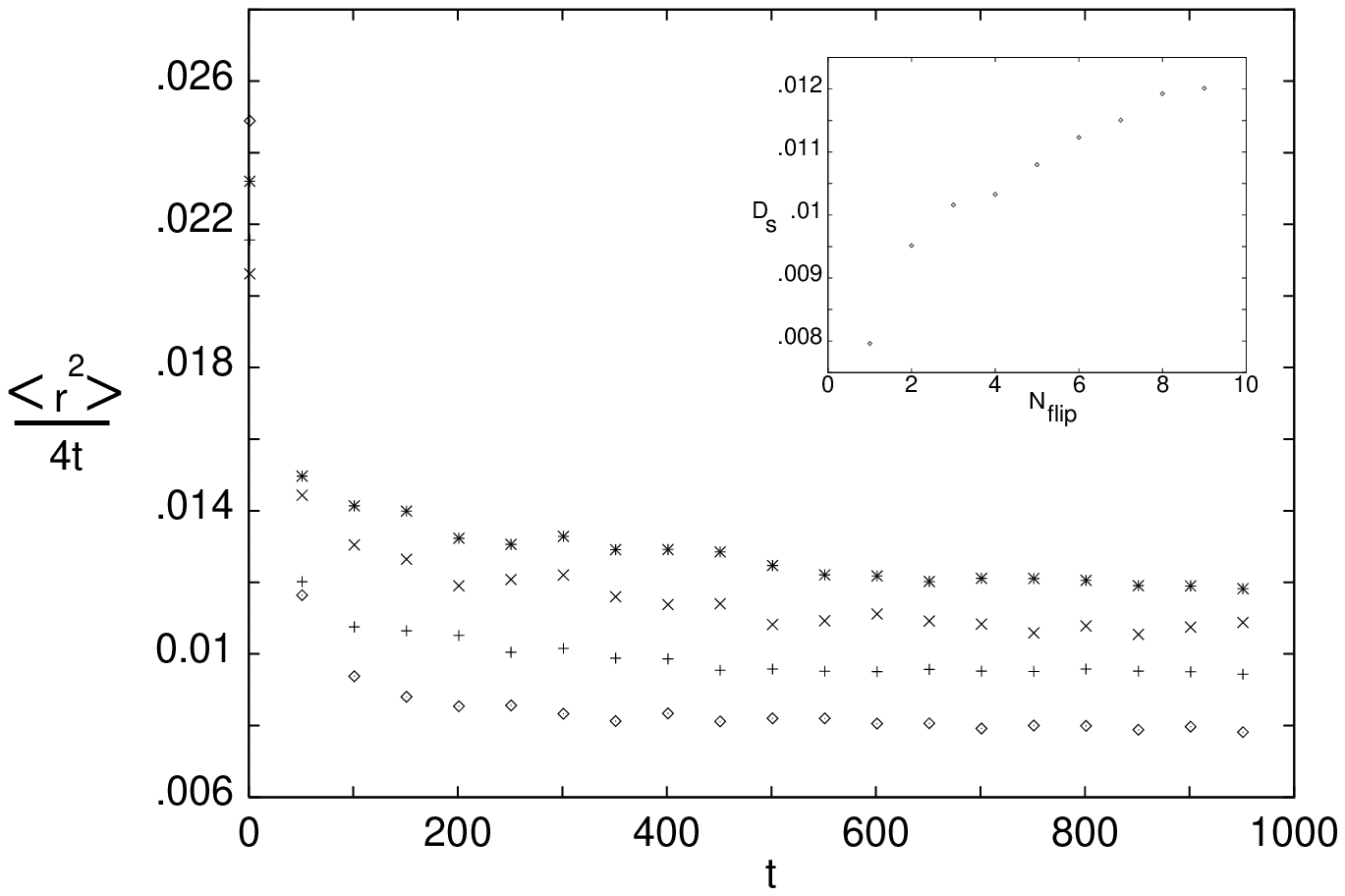}}{\vskip-1.8inFIG.~1. \
The single particle diffusion coefficient $D_{s}$ at different $N_{flip}$,
for $N_{flip} = N$, $2N$, $5N$ and $9N$ going from bottom to top, 
is given by the value of $\langle r^2 \rangle/4t$ at the plateau.  
The inset shows $D_{s}$ as a function of $N_{flip}$ (in units of $N$). 
The area fraction $N\pi a^2/A = 0.14$, is kept fixed.}

A crucial test of whether our MC algorithm correctly describes the hydrodynamic
behaviour of fluids, is the occurence of ``long-time tails'' in the measured 
velocity
autocorrelation function $Z(t) \equiv <{\bf u}_i(0) \cdot {\bf u}_i(t)>$
of dense (or highly viscous) fluids \cite{HANSEN}. We compute $Z(t)$ (averaged
over several initial conditions and particles) for different values of $N_{flip}$ and density. We find a convincing $t^{-1}$ tail (Fig.\ 2)
in the $Z(t)$ of high density (large $N$) and high viscosity (low $N_{flip} =
0.25N$) 
fluids. The inset of Fig.\ 2 shows the exponential decay (Enskog
result) of $Z(t)$ for lower viscosity ($N_{flip} = 10N$).
Note that the early time fall from $Z(0)=1$, does not show up, since
for the densities under consideration, several ``collisions'' have taken place within one MC time step. The inset also displays the $t^{-1}$ tail for
comparison. 
\vskip.5in
\myfigure{\epsfysize4in\epsfbox{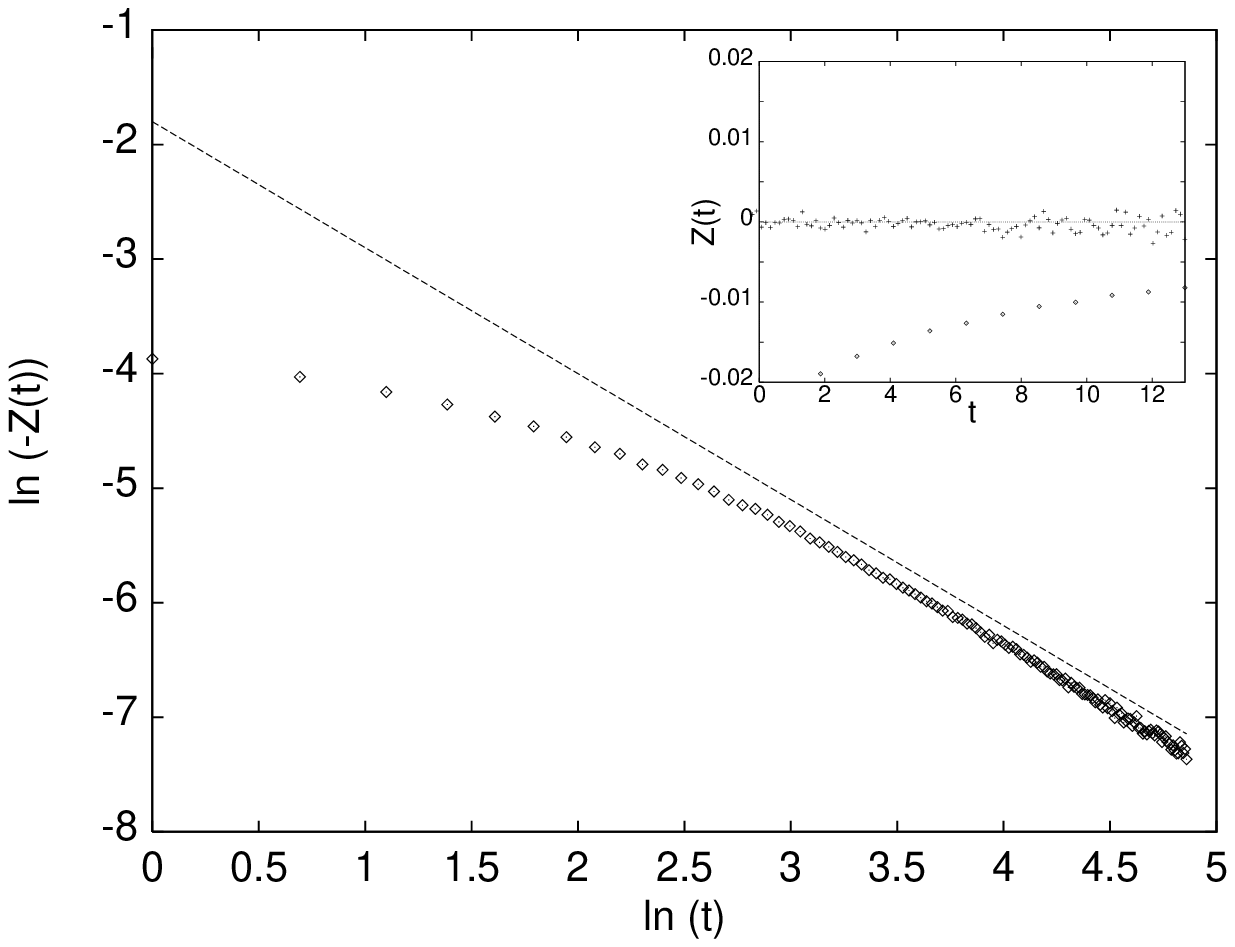}}{\vskip-2.2inFIG.~2. \
The velocity autocorrelation function $Z(t)$ for high density, high viscosity
fluids exhibits a clear $t^{-1}$ long-time tail.
Inset : $Z(t)$ decays to zero for low viscosity fluids (the $t^{-1}$ tail is displayed for comparison).}

Encouraged by this success, we study the dynamics of phase separation
of a binary fluid using our MC algorithm. Let us recount that when two immiscible fluids, like water+toluene, are cooled below their coexistence curve, they phase segregate 
into water-rich and toluene-rich regions, separated by sharp interfaces.
 At late times, the system enters a dynamical scaling regime \cite{REVIEW}, 
where the equal-time concentration correlation function behaves as $g(r,t) = g(r/t^{z})$.
The growth exponent $z$ is independent of microscopic details and depends on the existence of conservation laws. The scaling form defines a characteristic length scale $R(t)\sim t^{z}$, which measures
the typical distance between interfaces.

The phase separation dynamics of a 2-d binary fluid is described by a (conserved) concentration density $\phi({\bf r},t)$ coupled to a (conserved) momentum density ${\bf \pi}({\bf r},t)$. A variety of theoretical (based on 
dimensional analysis \cite{DIMEN,REVIEW}) and numerical (LS \cite{LANGSIM} and 
MD \cite{MDSIM}) techniques have been used to understand the late stage dynamics
of a 2-d binary fluid, but have unfortunately given rise to conflicting results.
The theoretical analysis and LS, contend that,
as in 3-d, the growth crosses over from a viscosity-dominated, $R \sim t$, 
to an inertia-dominated, $R \sim t^{2/3}$. On the other hand,
extensive MD simulations\cite{MDSIM} report a late time $R \sim t^{1/2}$ growth.

Our model 2-d binary fluid consists of two types of hard spheres, A and B (the total number of A ($N_A$) and B ($N_B$) beads 
are fixed, $N=N_A+N_B$). 
The local concentration $\phi_i$, defined as the difference in the densities of A and B, takes values
$+1$ (for A) and $-1$ (for B). The dynamics of $\phi$ is described by the 
usual Kawasaki exchange, which conserves $\sum_{i} \phi_i$. Thus locally
$\phi_i$ evolves by exchanging particles at vertices $i$ and $j$, where
$j$ is connected to $i$ by a tether, with a transition probability
$W(i \leftrightarrow j)=[1-\tanh(\Delta E/2k_BT)]/2$, where $\Delta E$ is the energy difference  between the final and the initial configuration.
The energy is calculated using the Ising hamiltonian
\begin{equation}
H_{ex} = -\,J\,\sum_{<ij>} \phi_{i} \phi_{j}\,\,,
\label{eq:ising}
\end{equation}
where the sum over $<ij>$ is over vertex pairs connected by a tether.

In addition to the two MC moves described above, each Monte Carlo sweep 
now includes $N_{ex}$ attempts at performing Kawasaki exchanges. Note that the 
Boltzmann weights associated with particle movements and the bond flips do not 
involve $H_{ex}$. It is clear that fluctuations in the local
exchange-energy are related to fluctuations in the local coordination number, which in turn are related to fluctuations in the local
velocity. In this way, the concentration variable $\phi$ gets coupled to the velocity $u$.

\myfigure{\epsfysize6in\epsfbox{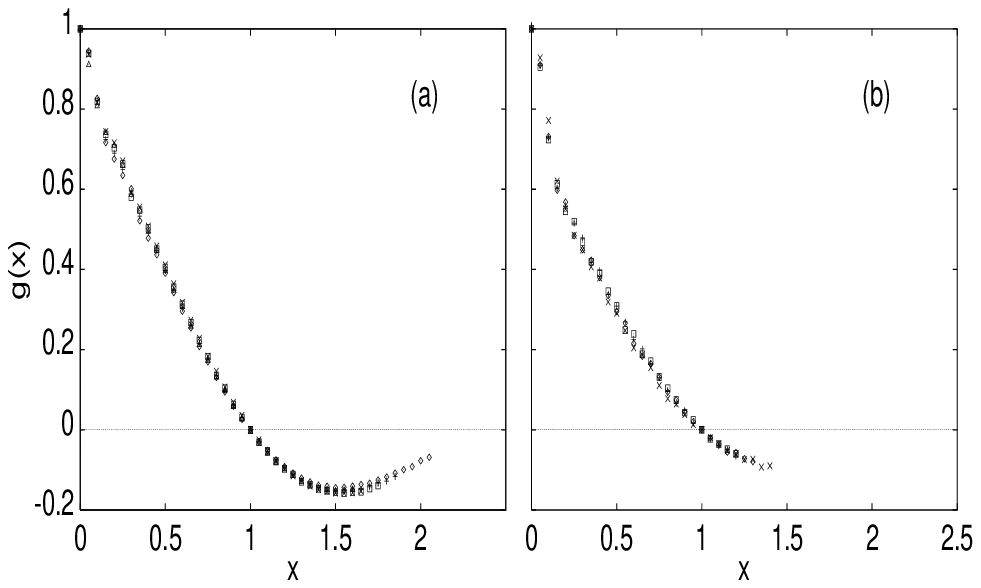}}{\vskip-3.8inFIG.~3. \
Scaling functions for the correlation function $g(r,t)$ at (a) high viscosity
($N_{flip}=30N$) and (b) low viscosity ($N_{flip}=100N$). Here $x = r/R(t)$ is the scaling variable.
}

At time $t=0$, our 2-d binary fluid is a homogeneous (50-50) mixture of A and B
in equilibrium at a high temperature $T$. 
We quench to below the critical temperature, $T < T_c \approx 2.25 J$
(for a triangular lattice). The homogeneous
state is unstable at this temperature, and evolves into a final phase separated state by the slow annealing of interfaces, conserving the order parameter $\phi$
during the process. A time sequence of typical configurations of $\phi$ (resembling those in Ref.\ \cite{REVIEW}), clearly exhibit statistical self-similarity at late times. As a 
quantitative measure, we compute the circularly averaged pair
correlation function 
$g(r,t)\equiv N^{-1} \sum_{\bf x} \langle \phi({\bf x}+{\bf r},t) \phi({\bf x},t) \rangle$, averaged 
over several realisations of the initial configurations of $\phi$
(it was sufficient to average over 10 realisations for good statistics).
 The domain size $R(t)$ is extracted from the 
first zero of the correlation function, $g(R(t),t) = 0$.
The correlation function, $g(r,t)$ exhibits dynamical scaling, $g(r,t)=g(r/R(t))$, at late times (Fig.\ 3), both for high viscosity 
$N_{flip}=30 N$ (Fig.\ 3a) and for low viscosity $N_{flip}=100 N$ (Fig.\ 3b) 
fluids. Our study seems to indicate that the scaling function depends on the viscosity of the fluid.

\myfigure{\epsfysize2.5in\epsfbox{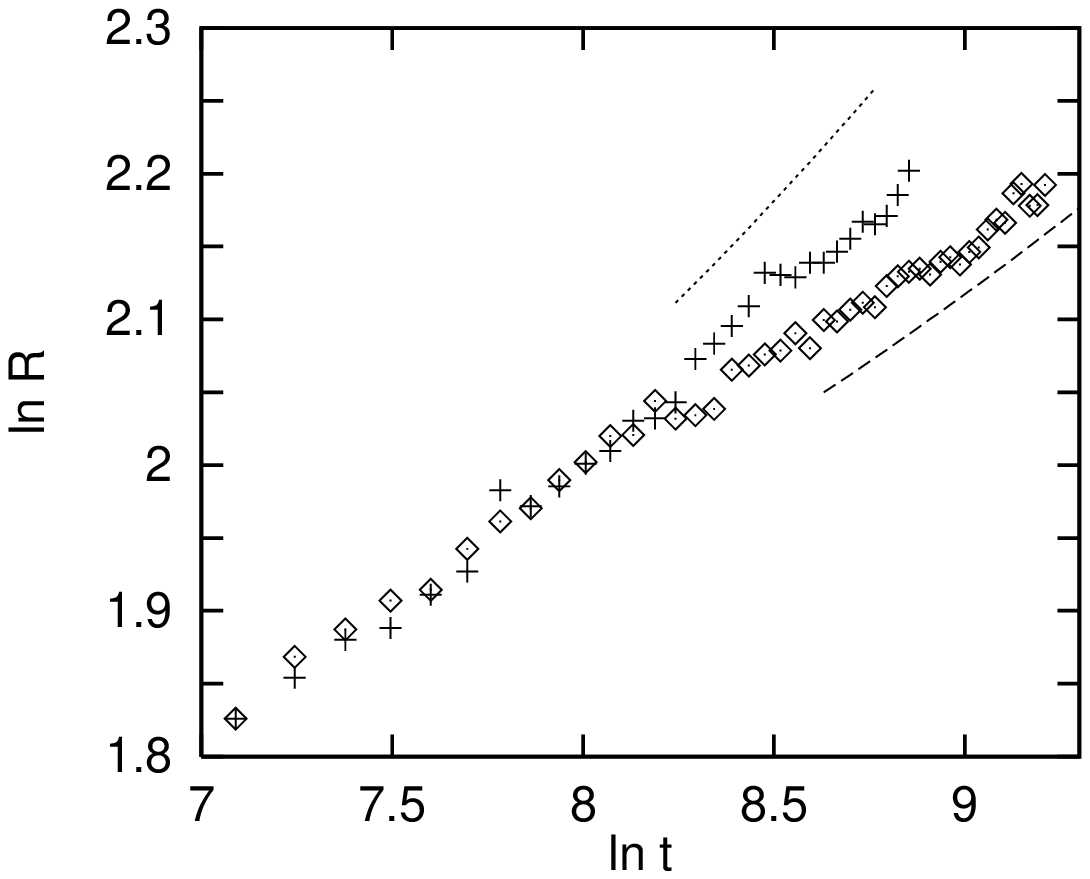}}{\vskip0inFIG.~4. \
Domain size (log-log plot) $R(t)$ scales as ($\diamond$) $t^{1/2}$  for high viscosity ($N_{flip}=30N$) and ($+$) $t^{2/3}$ for low viscosity ($N_{flip}=
100N$). The lines with slopes $1/2$ and $2/3$ are aids to the eye.
}

The domain size $R(t)$, determined both from the correlation function and
the interfacial energy density, is found to scale as $R(t) \sim t^{z}$.
As a preliminary check, we perform the usual Kawasaki dynamics in the 
absence of any bond flip and obtain a clean $t^{1/3}$ Lifshitz-Slyozov growth.
On introducing the flip moves that mimic fluidity, we see a dramatic crossover
to a growth influenced by hydrodynamics. For high viscosity
fluids, $N_{flip}=30N$, Fig.\ 4 shows a growth consistent with
$R \sim t^{1/2}$ at late times, before finite size effects become apparent.
A log-log plot for $N=7500$, shows a $z=0.48\pm 0.03$.
Simulations for larger system sizes and longer times, give 
the same value of $z$, indicating that this value of the exponent is robust
upto the late times we have investigated.
 This growth law is in agreement with MD simulations \cite{MDSIM}. 
On the other hand, for low viscosity fluids $N_{flip}=100N$, Fig.\ 4 shows a
late time exponent $z= 0.6\pm 0.05$, indicating a crossover 
 from the viscosity dominated $t^{1/2}$ growth to the inertia dominated $t^{2/3}$ growth. Thus we expect $R \sim t^{1/2}$
when $t \ll t_{\times}$, and  $R \sim t^{2/3}$ when $t \gg t_{\times}$,
where the crossover time $t_{\times} \sim \eta^{\alpha}$. 
Indeed, recent LB simulations \cite{LBSIM}, have seen the $t^{2/3}$ growth for low viscosity fluids. At higher viscosities they see an early time
$t^{1/3}$ growth, and claim a crossover to a later time $t^{2/3}$. 
{\it However their published data} \cite{LBSIM} 
{\it clearly indicates a crossover to} $t^{1/2}$. We claim that the LB results 
are consistent with our MC simulations.
As is clear from the above discussion, our
computational costs are low. This crossover from $z=1/2$ for high viscosities, to $z=2/3$ for low viscosities, can be understood theoretically, and we defer a discussion to another paper\cite{MADSUN}. 

In this Letter, we have succesfully implemented a Monte Carlo algorithm,
which describes the late-time dynamics of fluids in two dimensions.
This algorithm incorporates the momentum density of the fluids.
We have tested this code on a hard sphere fluid and have computed the
long-time tail of the velocity autocorrelation function. We have studied the phase separation dynamics of a binary fluid, and find that the domain size 
$R(t) \sim t^{1/2}$ for high viscosity, and crosses over to  
$R(t) \sim t^{2/3}$ for low viscosity fluids, with a crossover time depending
on the viscosity. This algorithm can be easily extended
to allow for an arbitrary pair potential $V(r)$. In a subsequent publication we
shall develop an analogous MC algorithm to study the late-time hydrodynamics
of fluids in three dimensions.

We have enjoyed discussions with Surajit Sengupta, Sriram Ramaswamy and
Rahul Pandit.

\end{document}